\documentclass[prc,preprint,showpacs,superscriptaddress,floatfix,amsmath,amssymb,nofootinbib]{revtex4}
\usepackage{graphicx}
\usepackage{color}
\usepackage{mathrsfs}

\newcommand{\beq}{\begin{equation}}
\newcommand{\eeq}{\end{equation}}
\newcommand{\beqn}{\begin{eqnarray}}
\newcommand{\eeqn}{\end{eqnarray}}

\newcommand{\Lb}{\left\{}
\newcommand{\Rb}{\right\}}

\newcommand{\ls}{\left[}
\newcommand{\rs}{\right]}
\newcommand{\lr}{\left|}
\newcommand{\rl}{\right|}
\newcommand{\re}{\nonumber\\}

\begin{document}

\title{Chiral bands for quasi-proton and quasi-neutron coupling with a triaxial rotor}

\author{S.Q. Zhang}\thanks{sqzhang@pku.edu.cn}

\affiliation{School of Physics, and MOE Key Laboratory of Heavy Ion
Physics, Peking University, Beijing 100871, China}
\affiliation{Institute of Theoretical Physics, Chinese Academy of
Sciences, Beijing, 100080, China}

\author{B. Qi}

\affiliation{School of Physics, and MOE Key Laboratory of Heavy Ion
Physics, Peking University, Beijing 100871, China}

\author{S.Y. Wang}

\affiliation{School of Physics, and MOE Key Laboratory of Heavy Ion
Physics, Peking University, Beijing 100871, China}

\author{J. Meng} \thanks{mengj@pku.edu.cn}

\affiliation{School of Physics, and MOE Key Laboratory of Heavy Ion
Physics, Peking University, Beijing 100871, China}
\affiliation{Institute of Theoretical Physics, Chinese Academy of
Sciences, Beijing, 100080, China}
 \affiliation{Center of Theoretical Nuclear Physics, National
 Laboratory of Heavy Ion Accelerator, Lanzhou 730000, China}

\date{\today}

\begin{abstract}

A particle rotor model (PRM) with a quasi-proton and a quasi-neutron
coupled with a triaxial rotor is developed and applied to study
chiral doublet bands with configurations of a $h_{11/2}$ proton and
a $h_{11/2}$ quasi-neutron. With pairing treated by the BCS
approximation, the present quasi-particle PRM is aimed at simulating
one proton and many neutron holes coupled with a triaxial rotor.
After a detailed analysis of the angular momentum orientations,
energy separation between the partner bands, and behavior of
electromagnetic transitions, for the first time we find aplanar
rotation or equivalently chiral geometry beyond the usual one proton
and one neutron hole coupled with a triaxial rotor.
\end{abstract}


\maketitle

\section{Introduction}

Since the pioneering work of Frauendorf and Meng~\cite{FM97}, the
phenomenon of chiral rotation in atomic nuclei has attracted
significant attention. Chirality in nuclei offers direct evidence
for the existence of stable triaxial nuclear shapes, in which there
are a few high-$j$ valence particles and a few high-$j$ valence
holes. For a triaxially deformed rotational nucleus, the collective
angular momentum favors alignment along the intermediate axis, which
in this case has the largest moment of inertia, while the angular
momentum vectors of the valence particles (holes) favor alignment
along the nuclear short (long) axis. The three mutually
perpendicular angular momenta can be arranged to form two systems
with opposite chirality, namely left- and right-handedness. They are
transformed into each other by the chiral operator which combines
time reversal and spatial rotation of 180$^\circ$, $\chi={\cal
TR}(\pi)$. The spontaneous breaking of chiral symmetry thus happens
in the body-fixed reference frame.  In the laboratory reference
frame, with the restoration of chiral symmetry due to quantum
tunneling, the so-called chiral doublet bands, a pair of separated
$\Delta I = 1$ bands (normally regarded as nearly degenerate) with
the same parity, are expected to be observed in triaxial nuclei.

Originally a pair of $\Delta I = 1$ bands found in $^{134}$Pr with
the $\pi h_{11/2}\otimes \nu h_{11/2}$
configuration~\cite{Petrache96}, has been reinterpreted in
Ref.~\cite{FM97} as a candidate for chiral doubling. Thereafter,
similar low-lying doublet bands were reported in $_{55}$Cs,
$_{57}$La, and $_{61}$Pm $N=75$ isotones of $^{134}$Pr, and an
island of chiral rotation was suggested in the $A \sim 130$ mass
region~\cite{Starosta01}. So far, candidate chiral doublet bands
have been proposed in a number of odd-odd nuclei in the $A \sim
130$~\cite{Starosta01,Koike01,Hecht01,Hartley01,Bark01, LiXF02,
Koike03, Rainovski03, Simons05, WangSY06a} and $A \sim 100$ mass
regions~\cite{Vaman04, Joshi04, Joshi05}. A few more candidates with
more than one valence-particle and hole were also reported in
odd-A~\cite{Zhu03, Alcantara04, Timar04, Timar06} and even-even
nuclei~\cite{Mergel02}.

On the theoretical side, chiral bands were first predicted in the
particle-rotor model (PRM) and tilted axis cranking (TAC) approach
for triaxially deformed nuclei~\cite{FM97}. Numerous efforts have
been devoted to the development of the PRM and TAC approaches.
Chiral rotation has been studied by the Strutinsky shell correction
TAC (SCTAC) method with a hybrid potential which combines the
spherical Woods-Saxon single-particle energies and the deformed part
of the Nilsson potential~\cite{Dimitrov00PRL,Dimitrov00PRC}.
Recently, chiral TAC solutions have also been found in $N=75$
isotones within the self-consistent Skyrme Hartree-Fock cranking
model~\cite{Olbratowski04,Olbratowski06}. The cranked relativistic
mean field (RMF) theory has been reported only in the contexts of
principle axis rotation~\cite{Koepf89,Afanasjev00} and planar
rotation~\cite{Madokoro00}. The generalization thereof for searching
for chiral solutions, i.e., the aplanar rotation, is still under
development. In Ref.~\cite{MengJ06}, the adiabatic and
configuration-fixed constrained triaxial RMF approaches were
developed to obtain the nuclear potential energy surface with the
triaxial degree of freedom and the existence of multiple chiral
doublets (M$\chi$D) was predicted for $A\sim 100$ mass region based
on their triaxial deformations and their corresponding proton and
neutron configurations. The advantage of the cranked mean field
approach to describe nuclear rotation bands is that it can be easily
extended to the multi-quasiparticle case. However, the usual
cranking approach is a semiclassical model, where the total angular
momentum is not a good quantum number, and the description of
quantum tunneling of chiral partners is beyond the mean field
approximation~\cite{MengJ06,PengJ03,Frauendorf01}.

In contrast, the PRM is a quantum mechanical model where total
angular momentum is a good quantum number.  The model describes the
system in the laboratory reference frame and yields directly the
energy splitting and tunneling between doublet bands. Chirality for
nuclei in $A\sim100$ and $A\sim130$ regions has been studied with
the particle-rotor model for certain particle-hole
configurations~\cite{PengJ03,PengJ03cpl}, or the
core-quasiparticle/core-particle-hole coupling
model~\cite{Koike03,Starosta02} following the
Kerman-Klein-D\"{o}nau-Frauendorf method~\cite{Klein00}. Selection
rules of electromagnetic transitions for chiral doublet bands have
been proposed based on a simple particle-hole-triaxial rotor
model~\cite{Koike04}.

Though various versions of PRM and TAC have been applied to study
chiral bands, the essential starting point for understanding their
properties is based on one particle and one hole coupled with a
rigid triaxial rotor. Based on this scenario, a set of observable
signatures have been suggested as fingerprints of chiral
bands~\cite{FM97,Koike02,Vaman04,Petrache06,WangSY06b}. Critical
analyses for the representative cases of candidate chiral bands,
$^{134}$Pr in $A\sim 130$~\cite{Petrache06}, and $^{104,106}$Rh in
$A\sim 100$~\cite{WangSY06b} have been carried out. It has been
found that these candidate chiral bands in $^{134}$Pr and $^{104}$Rh
do not agree with all of those expected for chiral bands, although
these candidates have been considered as the best examples of chiral
rotation in the $A \sim$ 130 and $A\sim$100 mass regions (due to
their extremely small level discrepancy between the doublet bands).
Lifetime measurements are essential for extracting the absolute
$B(M1)$ and $B(E2)$ transition probabilities, which are critical
experimental observables in addition to the level energies. Indeed,
this has stimulated experimental programs aimed at identifying
chiral doublet bands~\cite{Tonev06,Grodner06}.

On the other hand, one should bear in mind that these fingerprints
of chiral bands are obtained mostly by assuming one proton (neutron)
particle and one neutron (proton) hole sitting in a high-$j$ shell
coupled with a triaxial rotor with $\gamma = 30^\circ$. In a
realistic nucleus, it is more natural that there will be more than
one nucleon in a high $j$-shell, e.g., the candidate chiral doublet
bands reported for $N=75$ isotones with $Z=$ 55 ($^{130}$Cs), 57
($^{132}$La), 59 ($^{134}$Pr), 61 ($^{136}$Pm) and 63 ($^{138}$Eu),
and for $Z=55$ (Cs) isotopes with $N=$ 69, 71, 73, 75, and 77. The
Fermi energy of a proton (neutron) will undoubtedly change with $Z$
($N$) in these isotones (isotopes). Therefore it is interesting and
necessary to investigate the doublet bands with valence nucleons
sitting in the middle of a high $j$-shell, or alternatively
multi-particles sitting in a high $j$-shell. It is also important to
investigate the properties of doublet bands as functions of the
triaxial deformation degree of freedom.

To address these issues, in this paper a particle rotor model with a
quasi-proton and a quasi-neutron coupled with a triaxial rotor is
developed and applied to study chiral doublet bands with
configurations of a $h_{11/2}$ proton and a $h_{11/2}$
quasi-neutron.  With the pairing correlations taken into account by
the BCS approximation, the configuration of multi-particles sitting
in a high $j$-shell can be simulated by adjusting the neutron Fermi
energy. Note that in a former paper~\cite{WangSY07}, the present
model has been applied to the doublet bands of $^{126}$Cs, and good
agreement with the data available was obtained, which supports the
chiral interpretation of these doublet bands. Here the formalism is
given in detail and the properties of doublet bands calculated are
presented. The model is introduced in Sec. II. The properties of
doublet bands thus obtained, such as energy spectra, electromagnetic
transitions, and the orientation of angular momenta, are discussed
in Sec. III. Finally, a summary and conclusion is given in Sec. IV.

\section{Formalism}

{The particle rotor model~\cite{Bohr75} for triaxial deformed case
has been well used for the description of the odd-$A$ and odd-odd
nuclei~\cite{Meyer75,Ring81,Larsson78,Ragnarsson88,PengJ03}. Its
Hamiltonian for an odd-odd nucleus can be expressed as,
 \beq
 \label{eq:hamiltonian}
    H=H_\textrm{coll}+ H_\textrm{intr}^p+H_\textrm{intr}^n,
 \eeq
where $p$ and $n$ refer to protons and neutrons, respectively. The
collective Hamiltonian takes the form
 \beq
 H_\textrm{coll}
 = \sum_{i=1}^{3} \frac{\hat{R}_{i}^2}{2{\cal J}_i}
 = \sum_{i=1}^{3}
\frac{(\hat{I}_{i}-\hat{j}_{pi}-\hat{j}_{ni})^{2}}{2{\cal J}_i} ,
 \label{eq:hcoll}
 \eeq
where $\hat{R}_i, \hat{I}_{i}, \hat{j}_{pi}, \hat{j}_{ni}$
respectively denote the angular momentum operators for the core,
nucleus, as well as the valence proton and neutron. The moments of
inertia for irrotational flow are adopted, i.e., ${\cal J}_i = {\cal
J}\sin^2(\gamma - \displaystyle\frac{2\pi}{3}i)$ .}

{The intrinsic Hamiltonian for valence nucleons is
 \beq
 H_\textrm{intr}^{p (n)}
 = H_\textrm{sp}+H_\textrm{pair}
 = \sum_{\nu >0 }(\varepsilon_\nu-\lambda)
   (a_\nu^{+}a_\nu + a_{\overline{\nu}}^{+}a_{\overline{\nu}} )
 -\frac{\Delta}{2}\sum_{\nu >0}
 (a_{\nu}^{+}a_{\overline\nu}^{+}+a_{\overline{\nu}}{a_\nu}),
 \eeq
where $\lambda$ denotes the Fermi energy, $\Delta$ the pairing gap
parameter, and $|\overline{\nu}\rangle$ the time-reversal state of
$|{\nu}\rangle$. The single particle energy $\varepsilon_\nu$ is
obtained by the diagonalization of the Hamiltonian $H_\textrm{sp}$.
 Similar as in Ref.~\cite{PengJ03}, for a single-$j$ shell, one has
 \beq\label{eq:hsp}
 H_\textrm{sp}=\pm \frac{1}{2}C
    \Lb\cos\gamma(j_3^2-\frac{j(j+1)}{3})
    + \frac{\sin\gamma}{2\sqrt{3}}(j_+^2+j_-^2)\Rb,
 \eeq
where the plus sign refers to a particle, the minus to a hole, and
the coefficient $C$ is proportional to the quadrupole deformation
$\beta$~\cite{Meyer75,PengJ03}. The single particle states are thus
written as
 \beq
\label{eq:spwf}
    {a}^{+}_{\nu}| 0 \rangle
    =\sum_{\Omega}c_{\Omega}^{(\nu)}\psi^{j}_{\Omega},
    ~~~~~
  {a}^{+}_{\overline{\nu}}| 0 \rangle
 = \sum_{\Omega}(-1)^{j-\Omega}c_{\Omega}^{(\nu)}\psi^{j}_{-\Omega},
 \eeq
where $\Omega$ is the projection of the single-particle angular
momentum ${\hat j}$ along the 3-axis and can be restricted to the
values $\cdots, -7/2, -3/2, +1/2, +5/2, \cdots$ due to time-reversal
degeneracy.~\cite{Larsson78,Ragnarsson88}}

To obtain the PRM solutions, the total
Hamiltonian~(\ref{eq:hamiltonian}) must be diagonalized in a
complete basis space, which couples the rotation of the inert core
with the intrinsic wave functions of valence nucleons. When pairing
correlations are neglected, one can construct the so-called strong
coupling basis as
  \beqn\label{eq:base1}
  |IMK\nu_p\nu_n\rangle &=&\sqrt{\frac{1}{2}}
  \sqrt{\frac{2I+1}{8\pi^2}}\ls
  D_{M,K}^{I} a^{+}_{\nu_p}
  a^{+}_{\nu_n} |0\rangle
   +(-1)^{I-K}D_{M,-K}^{I} a^{+}_{\overline{\nu}_p}
  a^{+}_{\overline{\nu}_n} |0\rangle
  \rs\nonumber\\
  &=&  \sqrt{\frac{2I+1}{16\pi^2}}
  \sum_{\Omega_p}\sum_{\Omega_n}c_{\Omega_p}^{(\nu_p)}c_{\Omega_n}^{(\nu_n)}
 \ls
  D_{M,K}^{I}\psi^{j_p}_{\Omega_p}\psi^{j_n}_{\Omega_n}
  +(-1)^{I-j_p-j_n}D_{M,-K}^{I}\psi^{j_p}_{-\Omega_p}\psi^{j_n}_{-\Omega_n}\rs
 \nonumber \\
{\rm for~} K&=&\pm 1, \pm3, \pm5\cdots,
 \eeqn
 \beqn\label{eq:base2}
  |IMK\nu_{p}\overline{\nu}_n\rangle &=&\sqrt{\frac{1}{2}}
  \sqrt{\frac{2I+1}{8\pi^2}}\ls
  D_{M,K}^{I} a^{+}_{\nu_p}
  a^{+}_{\overline{\nu}_n}| 0 \rangle
   +(-1)^{I-K}D_{M,-K}^{I} a^{+}_{\overline{\nu}_p}
  a^{+}_{\nu_n}| 0 \rangle
  \rs\nonumber\\
  &=&\sqrt{\frac{2I+1}{16\pi^2}}
  \sum_{\Omega_p}\sum_{\Omega_n}
  c_{\Omega_p}^{(\nu_p)}c_{\Omega_n}^{(\nu_n)}(-1)^{j_n-\Omega_n}
  \ls D_{M,K}^{I}\psi^{j_p}_{\Omega_p}\psi^{j_n}_{-\Omega_n}
     +(-1)^{I-j_p-j_n}D_{M,-K}^{I}\psi^{j_p}_{-\Omega_p}\psi^{j_n}_{\Omega_n}\rs
 \nonumber\\
\textrm{for~} K&=&0, \pm2, \pm4\cdots.
    \eeqn
The restriction on values of $K$ is due to the fact that the basis
states are symmetrized under the point group $D_2$, which leads to
$K-\Omega_p-\Omega_n$ in Eq.~(\ref{eq:base1}) and
$K-\Omega_p+\Omega_n$ in Eq.~(\ref{eq:base2}) being an even
integer~\cite{Larsson78}. The matrix elements of Hamiltonian
(\ref{eq:hcoll}) and (\ref{eq:hsp}) can be evaluated in the basis
(\ref{eq:base1}) and (\ref{eq:base2}), and then diagonalization
gives eigenenergies and eigenstates for the PRM Hamiltonian. For a
certain spin $I$, the dimension of the basis space will be
$(1/4)(2I+1)(2j_p+1)(2j_n+1)$.

To include pairing effects in the PRM, one should replace the single
particle state $a^{+}_{\nu}| 0\rangle$ in the basis
states~(\ref{eq:base1}) and (\ref{eq:base2}) with the BCS
quasiparticle state $\alpha^{+}_{\nu}|\tilde{0}\rangle$ to obtain a
new expansion basis, where $|\tilde{0}\rangle$ is the BCS vacuum
state. The quasiparticle operators $\alpha_{\nu}^{+}$ are given by
 \beq
 \left(
\begin{array}{c}
  \alpha_{\nu}^{+} \\
  \alpha_{\overline{\nu}} \\
\end{array}
\right) =\left(\begin{array}{cc}
  u_\nu & -v_\nu \\
  v_\nu & u_\nu \\
\end{array} \right)~~
\left(
\begin{array}{c}
  a_{\nu}^{+} \\
  a_{\overline{\nu}} \\
\end{array}\right),
 \eeq
where $u_\nu^2 + v_\nu^2 = 1$. In this new basis, the wave functions
of PRM Hamiltonian are written as
 \beq
 |IM \rangle =
 \sum_{K,\nu_p,\nu_n} \left(C_{\nu_p\nu_n}^{IK}
 |IMK\nu_p\nu_n\rangle + C_{\nu_p\overline{\nu}_n}^{IK}
 |IMK\nu_p\overline{\nu}_n\rangle \right),
 \eeq
in which $\nu_p$ and $\nu_n$ represent the quasiparticle states
$\alpha^{+}_{\nu_p}|\tilde{0}\rangle$ and
$\alpha^{+}_{\nu_n}|\tilde{0}\rangle$ instead. Furthermore,
single-particle energies $\varepsilon_{\nu }$ should be replaced by
quasiparticle energies
 $
 \varepsilon'_{\nu}= \sqrt{(\varepsilon_{\nu} -
 \lambda)^2+ \Delta^2}.
 $
 The total Hamiltonian then becomes:
 \beq H = H_{coll} +
 \sum_{\nu_p}\varepsilon'_{\nu_p}(\alpha_{\nu_p}^{+}\alpha_{\nu_p}
    + \alpha_{\overline{\nu}_p}^{+}\alpha_{\overline{\nu}_p})
+\sum_{\nu_n}\varepsilon'_{\nu_n}(\alpha_{\nu_n}^{+}\alpha_{\nu_n}
    + \alpha_{\overline{\nu}_n}^{+}\alpha_{\overline{\nu}_n}).
 \eeq
To construct the matrix of the above Hamiltonian, in comparison with
the case excluding pairing, each single-particle matrix element
needs to be multiplied by a pairing factor
$u_{\mu}u_{\nu}+v_{\mu}v_{\nu}$~\cite{Meyer75,Ragnarsson88}. The
occupation factor $v_\nu$ of the state $\nu$ is given by
 \beqn
v^2_\nu
  =\frac{1}{2}
   \left[1-\frac{\varepsilon_\nu-\lambda}{\varepsilon'_{\nu}}\right].
 \eeqn

The reduced electromagnetic transition probabilities are defined
as~\cite{Bohr75}
 \beq B(\sigma\lambda,
I\rightarrow I')=\sum_{\mu M'} |\langle I' M'|
 {\mathscr M}^\sigma_{\lambda\mu}|I M \rangle |^2,
 \eeq
where $\sigma$ denotes either $E$ or $M$ for electric and magnetic
transitions, respectively, $\lambda$ is the rank of transition
operator, and ${\mathscr M}^\sigma_{\lambda\mu}$ is the
electromagnetic transition operator.

For electric quadrupole (E2) processes, the corresponding transition
operator is generally taken as
 \beq
 {\mathscr M}(E2, \mu) = \int \rho_e ({\vec r}) r^2
     Y_{2\mu}(\theta,\phi) d\tau,
 \eeq
which is proportional to the electric quadrupole tensor operator
$\hat{Q}_{2\mu}$ with a factor $\sqrt{5/16\pi}$~. The quadrupole
moments in the laboratory frame ($\hat{Q}_{2\mu}$) and the intrinsic
system ($\hat{Q}'_{2\mu}$) are connected by the relation
 \beq
 \hat{Q}_{2\mu} = {\cal D}_{\mu 0}^{2*}\hat{Q}'_{20}
   + ({\cal D}_{\mu 2}^{2*}+{\cal D}_{\mu -2}^{2*})\hat{Q}'_{22}.
 \label{eq:Q}
 \eeq
{For stretched E2 transitions, one has
 \beqn
 &&B(E2,I\alpha\rightarrow I'\alpha)\re
 &=&
 \frac{5}{16\pi}Q_0^2\lr{
 \sum^{KK'}_{\nu_{p}\nu_{n}}
 C_{\nu_{p}\nu_{n}}^{IK}
 C_{\nu_{p}\nu_{n}}^{I'K'}\ls{\langle IK20|I'K'\rangle\cos\gamma
 +\frac{\sin\gamma}{\sqrt{2}} \bigl( \langle IK22|I'K'\rangle
 + \langle IK2-2|I' K'\rangle \bigr) }\rs}\rl^2 \re
 && + {\rm Term2},
 \label{eq:Be2a}
 \eeqn
where $Q_0$ is the intrinsic charge quadrupole moment and the
``\textrm{Term2}" term is the same as the first term but with the
replacement ($\nu_{n}\rightarrow \overline{\nu}_{n}$).}

For M1 transitions, the magnetic dipole transition operator can be
written as
 \beq
 \mathscr{M} (M1,\mu) = \sqrt{\frac{3}{4\pi}} \frac{e\hbar}{2Mc}
 [(g_p-g_R)\hat{j}_{p\mu}+(g_n-g_R)\hat{j}_{n\mu}],
 \eeq
where $g_p$, $g_n$, $g_R$ are respectively the effective
gyromagnetic ratios for valence proton, valence neutron and the
collective core, and $\hat{j}_{\mu}$ denotes the spherical tensor in
the laboratory frame.
 {The M1 reduce transition
 probability $B(M1)$
is expressed as
 \beqn
 && B(M1, I\alpha\rightarrow I'\alpha) \re
 &=& \frac{3}{4\pi} \biggl| \sum_{\mu KK'}C_{\nu_{p}\nu_{n}}^{IK}
C_{\nu'_{p}\nu'_{n}}^{I'K'}
 \sum_{\Omega'_p\Omega'_n}
 c_{\Omega'_p}^{(\nu'_{p})}c_{\Omega'_n}^{(\nu'_{n})}
 \sum_{\Omega_p\Omega_n}
 c_{\Omega_p}^{(\nu_{p})}c_{\Omega_n}^{(\nu_{n})}
 \re
&&
 \Bigl\{\langle{IK1\mu|I'K'}\rangle
 \langle{\Omega'_p \Omega'_n
 |\hat{T}_\mu|
 {\Omega}_p {\Omega}_n} \rangle
 + (-1)^{I-j_p-j_n}\langle{I -K 1 \mu|I'K'}\rangle
 \langle{\Omega'_p\Omega'_n
 |\hat{T}_\mu|
 -{\Omega}_p-{\Omega}_n}\rangle\Bigr\}\biggr|^2 \re
 && + \textrm{Term2} + \textrm{Term3} + \textrm{Term4},
 \label{eq:BM1}
 \eeqn
where terms ``\textrm{Term2}", ``\textrm{Term3}", ``\textrm{Term4}"
are the same as the first term but with the replacement
($\nu_{n}\rightarrow \overline{\nu}_{n}$), ($\nu'_{n}\rightarrow
\overline{\nu}'_{n}$), and ($\nu_{n}\rightarrow \overline{\nu}_{n}$,
$\nu'_{n}\rightarrow \overline{\nu}'_{n}$), respectively. The
operator $\hat{T}_\mu$ in Eq.(\ref{eq:BM1}) is given by
 \beq
\hat{T}_\mu=f(p)(g_p-g_R) \hat{j}_{p\mu} + f(n)(g_n-g_R)
\hat{j}_{n\mu},
 \eeq
with $f(p)$ and $f(n)$ the pairing factors $u u'+v v'$ for proton
and neutron, and $j_{\mu}$ the rank-1 spherical tensor in the
body-fixed reference frame.}
\section{Results and Discussion}

\subsection{Single particle states in the single-$j$ model}

\begin{center}
 \begin{figure*}[h!]
  \centering
  \includegraphics[width=10cm]{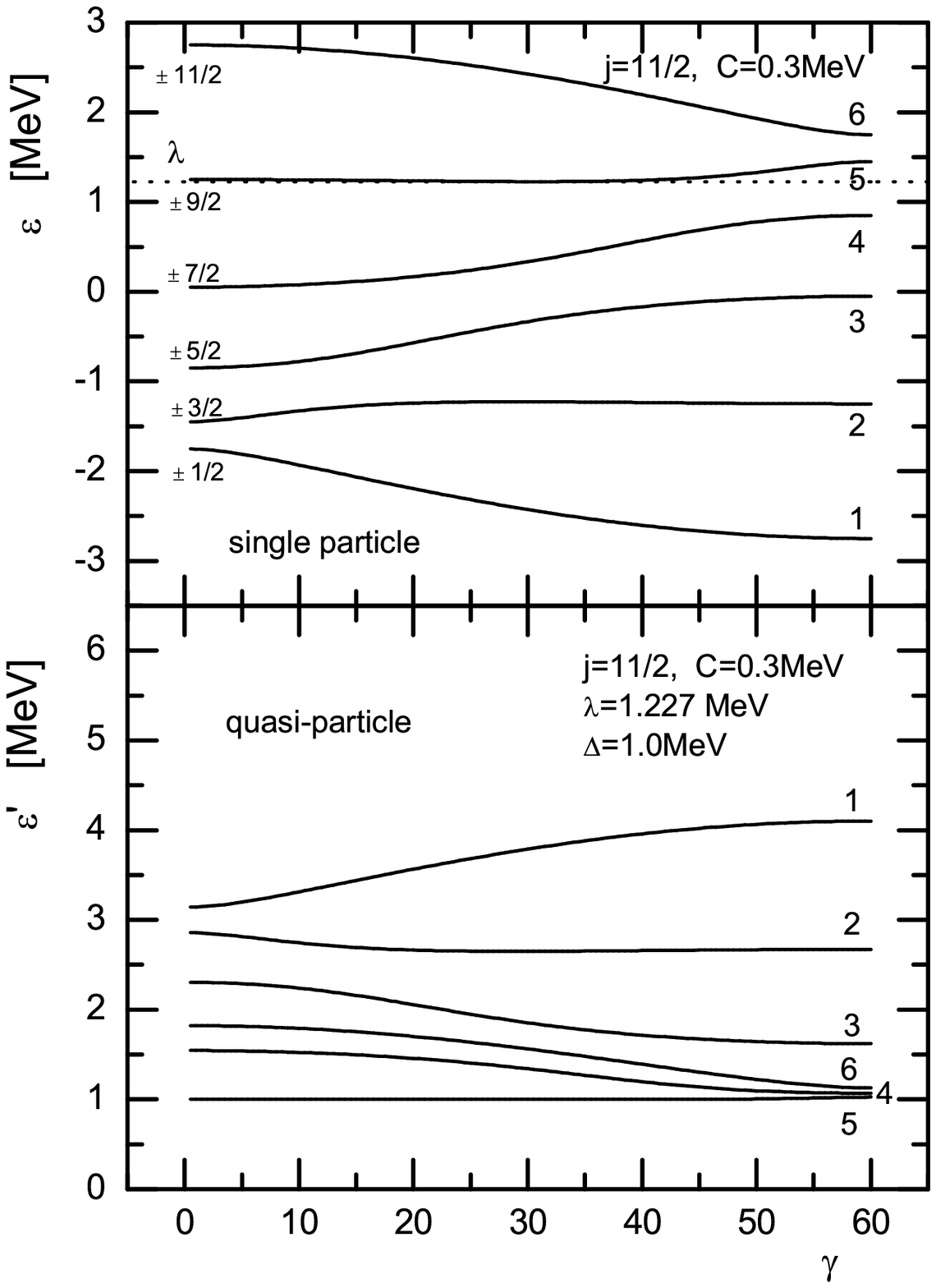}
   \caption{Upper panel: The single particle energy $\varepsilon$ with
   single-$j$ Hamiltonian in Eq.(\ref{eq:hsp}) ($j=11/2, C=0.3$ MeV)
   as a function of $\gamma$ deformation. The six degenerate levels are
   respectively indicated by 1, 2, $\cdots$, 6, as well as
   the corresponding third angular momentum components $\pm 1/2$, $\pm 3/2$,
   $\cdots$, $\pm 11/2$ at $\gamma = 0^\circ$ (which is good quantum
   number only for $\gamma = 0^\circ$). The dashed line indicates
   the Fermi energy $\lambda$, which is used to obtain the quasiparticle
   energy $\varepsilon'$ in the lower panel.
   Lower panel:  Quasiparticle energy $\varepsilon'$ for the same parameters
   as a function of $\gamma$ deformation. The pairing parameters
   are $\lambda =1.227$ MeV, $\Delta =$ 1 MeV. Each level (1, 2, $\cdots$, 6)
   corresponds to that with the same number in the upper panel.
   }
   \label{fig:spenergy}
\end{figure*}
\end{center}

For the intrinsic Hamiltonian of valence nucleons, we apply the
simple single-$j$ model, which is a good approximation for high-$j$
intruder orbitals~\cite{Bohr75}. The single particle energy
$\varepsilon$ corresponding to the Hamiltonian in Eq.(\ref{eq:hsp})
with $1h_{11/2}$ $j$-shell and $C = 0.3$ MeV is plotted in the upper
panel of Fig.~\ref{fig:spenergy} as a function of the $\gamma$
deformation of the deformed well. This $C = 0.3$ MeV corresponds to
a quadrupole deformation of $\beta\sim0.28$ for the $1h_{11/2}$
subshell in the $A\sim130$ mass region. When $\gamma=0^\circ$, i.e.,
axial symmetrical case, there are six discrete states with good
quantum number $\Omega$ ($\pm 1/2$, $\pm 3/2$, $\cdots$, $\pm
11/2$). These states are indexed by $\nu$ ($\nu$=1, 2, $\cdots$, 6),
and the corresponding energies are denoted by $\varepsilon_\nu$.
When axial symmetry is broken, $\Omega$ is not a good quantum
number, and each single particle state $\nu$ is then a superposition
of eigenstates of $(j^2, j_3)$ as in Eq.~(\ref{eq:spwf}), and
changes smoothly with $\gamma$. It can be clearly seen that for a
$h_{11/2}$ particle, a lower energy is obtained for
$\gamma=60^\circ$, i.e., an oblate shape is preferred, while for a
hole a prolate shape is preferred. Particularly, for a nucleus with
a $\pi h_{11/2}\otimes \nu h^{-1}_{11/2}$ configuration, the sum of
single particle energies will be fairly $\gamma$ soft with a minimum
around $\gamma=30^\circ$, and the $\gamma$ degree of freedom will
play an important role. Note that the single particle energies for
levels 2 and 5 are nearly $\gamma$ independent.

With pairing taken into account by the BCS calculation, the
quasiparticle energy $\varepsilon'$ with $\lambda =1.227$ MeV and
$\Delta =$ 1 MeV  is given in the lower panel of
Fig.~\ref{fig:spenergy}.  The Fermi energy $\lambda$ is very close
to $\varepsilon_5$, which is shown by a dashed line in the upper
panel. The label for each level follows the corresponding one in the
upper panel. Since the Fermi energy is $\lambda \simeq
\varepsilon_5$, the state $\varepsilon_5$ is now the lowest
quasiparticle state located at $\sim 1$ MeV due to the pairing gap
$\Delta$. Another feature is that the quasiparticle energy
$\varepsilon'_\nu$ becomes more $\gamma$ soft than the corresponding
single particle energy $\varepsilon_\nu$ due to pairing.


\subsection{Energy spectra}

In the present PRM, if $\lambda_n=\varepsilon_6$ and $\Delta_n=0$
for neutron, and $\lambda_p=\varepsilon_1$ and $\Delta_p=0$ for
proton, the model discussed here is equivalent to the model in
Refs.~\cite{FM97, PengJ03, Koike04}. In the following calculation,
$\lambda_p=\varepsilon_1$ and $\Delta_p=0$ are fixed for the proton,
i.e., a pure $h_{11/2}$ particle proton, while $\lambda_n$ for the
neutron changes from the bottom to the top of the $h_{11/2}$ shell.
The coefficient $C=0.3$ MeV, which corresponds to a quadrupole
deformation of $\beta\sim0.28$ for the $A\sim130$ mass region, and
the moment of inertia is ${\cal J}=30$ MeV$^{-1}$. For the
electromagnetic transition probabilities, the intrinsic charge
quadrupole momentum $Q_0$ takes a value of $3.5$ eb, and the
$g$-factors $g_p-g_R=0.7$ and $g_n-g_R=-0.6$ are adopted
respectively.

Firstly we investigate the behavior of doublet bands for a nucleus
at the deformation $\gamma=30^\circ$ in which the best chirality of
nuclear rotation is expected~\cite{FM97}. It should be noted that
for an asymmetric configuration $\pi g_{9/2}^{-1}\otimes\nu
h_{11/2}$, the best chirality occurs at a deformation
$\gamma=27^\circ$ in Ref.~\cite{PengJ03}.

\begin{center}
 \begin{figure*}[h!]
  \centering
  \includegraphics[width=8cm]{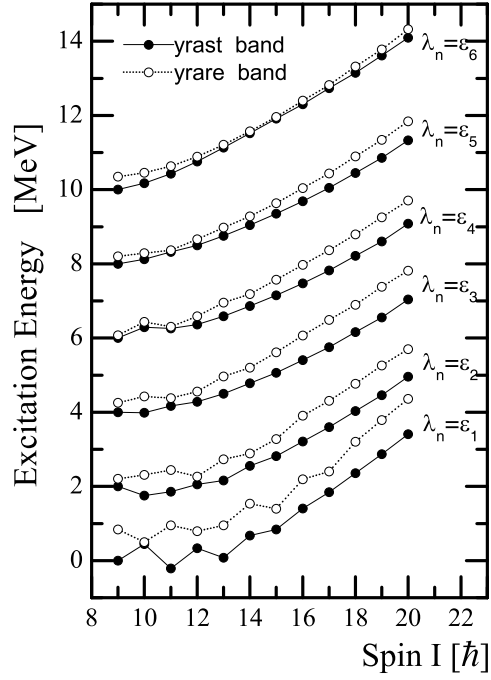}
   \caption{Calculated rotational spectra for the yrast (solid circles)
   and yrare (open circles) bands for
   the configuration $\pi h_{11/2}\otimes\nu h_{11/2}$ with $C=0.3$
   MeV, ${\cal J}=30$ MeV$^{-1}$, and $\gamma=30^\circ$. In the
   calculations, the odd proton is fixed to be a pure $h_{11/2}$
   particle, while the odd neutron is treated as a BCS quasiparticle
   with $\lambda_n=\varepsilon_1, \varepsilon_2, \cdots, \varepsilon_6$,
   respectively, and $\Delta =$ 1 MeV. The $I=9$ state energies of the yrast
   bands, assumed to be $0$ MeV,  are separated by 2.0 MeV for display.
   }
   \label{fig:spectra}
\end{figure*}
\end{center}

The calculated rotational spectra for the yrast and yrare
bands\footnote{In the paper, the yrast band denotes the rotational
band which connects the lowest energies with given spins $I$
obtained from the present PRM calculations, while the yrare band
correspondingly connects the second lowest energies.} with the
configuration $\pi h_{11/2}\otimes\nu h_{11/2}$ for $C=0.3$ MeV and
${\cal J}=30$ MeV$^{-1}$, are plotted in Fig.~\ref{fig:spectra}. In
the calculations, the odd proton is fixed to be a pure $h_{11/2}$
particle, while the odd neutron is treated as a BCS quasiparticle
with $\Delta =$ 1 MeV and $\lambda_n=\varepsilon_1, \varepsilon_2,
\cdots, \varepsilon_6$, respectively. The $I=9$ state energies of
the yrast bands are taken as reference points and are separated by
2.0 MeV for display. 

From Fig.~\ref{fig:spectra}, the energy difference between the yrare
and yrast bands increases from $\lambda_n=\varepsilon_6$ to
$\varepsilon_1$. For $\lambda_n=\varepsilon_6$, two nearly
degenerate bands can be clearly seen, especially for the spin
interval $13\leq I \leq 17$, and the energy difference between the
doublet bands is below 100 keV. This is the classical case where the
chiral concept was proposed~\cite{FM97}. When $\lambda_n=
\varepsilon_5$, the two bands are nearly degenerate with a constant
energy separation of $\sim 200$ keV for the spin interval $11\leq I
\leq 15$ and a gradually increasing energy separation for higher
spin. For $\lambda_n=\varepsilon_4$ and $\varepsilon_3$, the spectra
present very similar behavior: (1) at the low spin region $I<14$, a
slight odd-even staggering with opposite phase can be seen for yrast
and yrare bands; (2) only at low spins ($I=9, 11, 12$), the energy
difference of the yrast and yrare states is smaller than 250 keV;
(3) for spin $I\geq14$, the energy differences between yrast and
yrare states increase with $I$, e.g., $\sim 400$ keV at $I=14$ and
$\sim 700$ keV at $I=20$.  For $\lambda_n=\varepsilon_1$, odd-even
staggering becomes more obvious and the two bands are separated by
an average energy of $\sim 800$ keV. The case of
$\lambda_n=\varepsilon_2$ is in between that of $\varepsilon_3$ and
that of $\varepsilon_1$.

\begin{center}
 \begin{figure*}[h!]
  \centering
  \includegraphics[width=8cm]{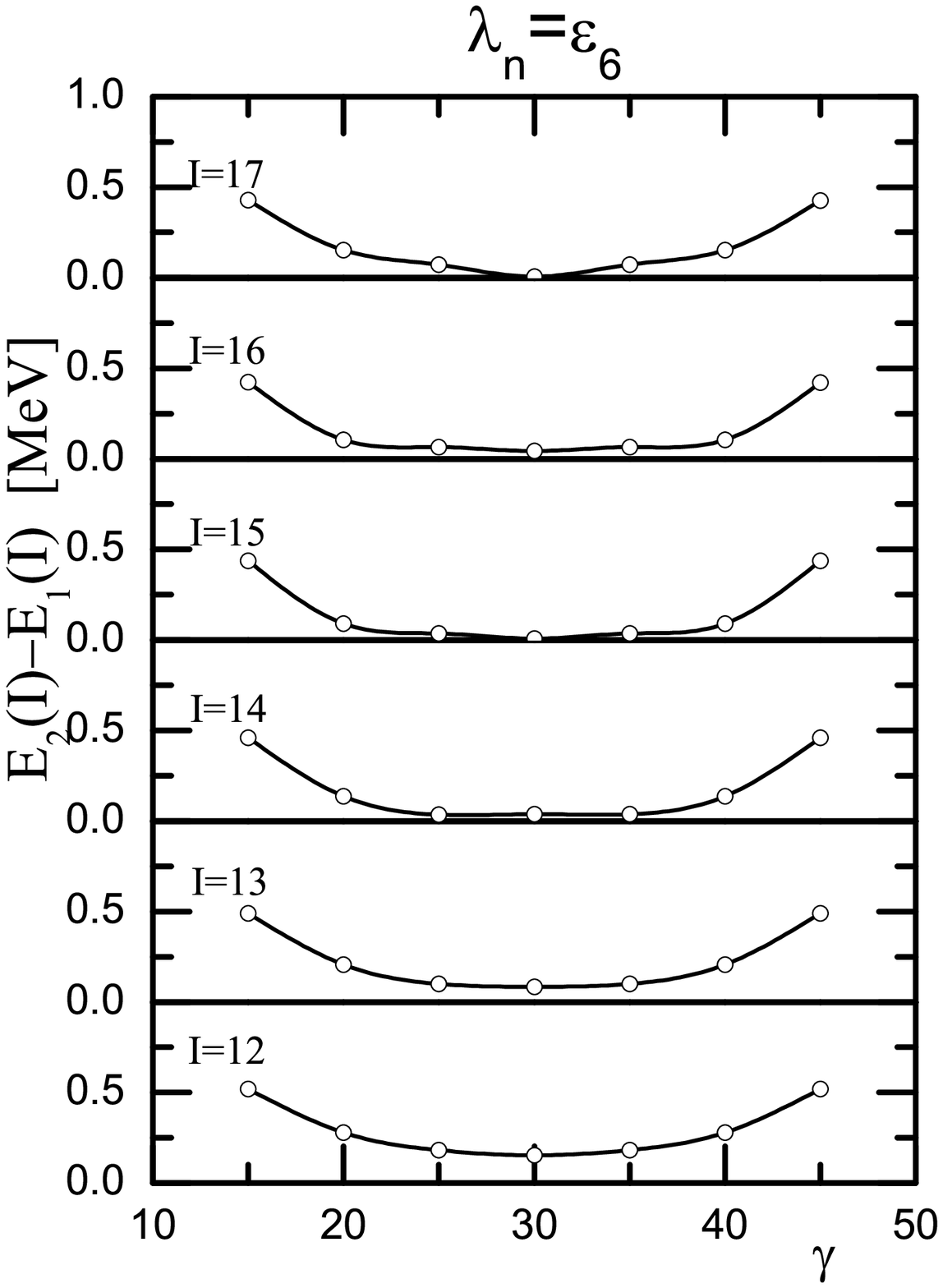}
  \includegraphics[width=8cm]{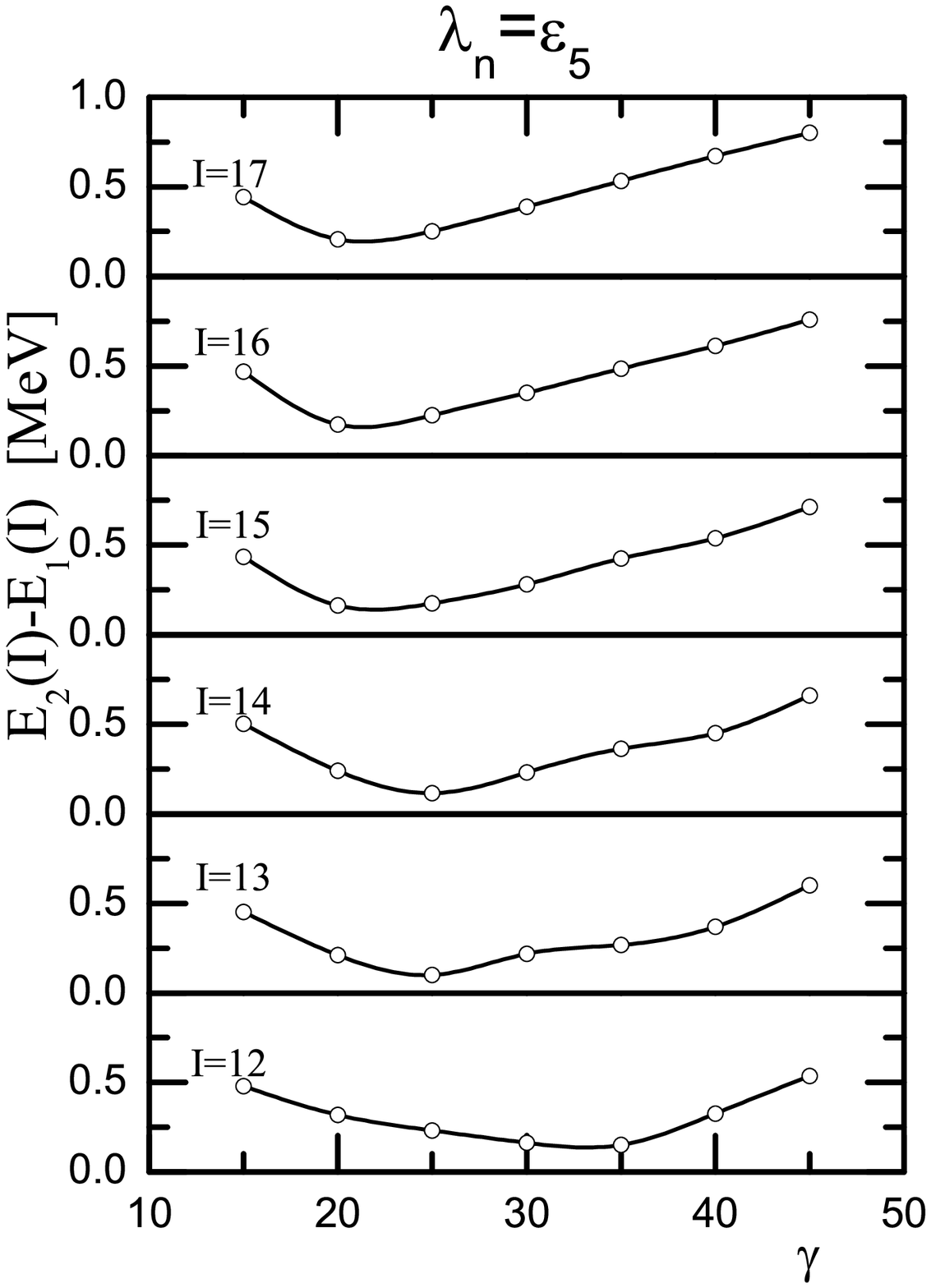}
   \caption{Calculated energy difference $E_2(I)-E_1(I)$ between
   yrare and yrast bands at $I=12, 13, \cdots, 17$ as a function
   of $\gamma$ deformation. In the calculations, $C=0.3$
   MeV, ${\cal J}=30$ MeV$^{-1}$ and the odd proton
   is fixed to be a pure $h_{11/2}$ particle, while the odd neutron is treated
   as a BCS quasiparticle with $\Delta =$ 1 MeV,
   and $\lambda_n=\varepsilon_6$ (Left panel), $\varepsilon_5$ (Right
   panel).
   }
   \label{fig:specgam}
\end{figure*}
\end{center}

The calculated energy difference $E_2(I)-E_1(I)$ between yrare and
yrast bands at spins $I=12, 13, \cdots, 17$ as a function of
$\gamma$ deformation is plotted in Fig.~\ref{fig:specgam}. The left
panel displays the results for a pure $h_{11/2}$ proton particle and
a pure $h_{11/2}$ neutron hole ($\lambda_n=\varepsilon_6$,
$\Delta=0$) configuration. A symmetric $E_2(I)-E_1(I)$ curve about
$\gamma=30^\circ$ can be seen, which in turn is associated with the
symmetries of Hamiltonians with respect to $\gamma=30^\circ$. If we
use $\Delta=1$ MeV instead for neutrons, the symmetry will not
strictly hold any more. In detail, the smallest energy difference
($< 200$ keV) takes place at $\gamma=30^\circ$ for all the shown
spins, and particularly at spins $I=15, 17$, very good degeneracy is
obtained, namely the energy
differences are 7.2 keV and 4.1 keV, respectively. 
The energy difference
increases when the $\gamma$ degree deviates from $30^\circ$, and
presents a parabola-like curve. At $\gamma=20^\circ$ and $40^\circ$,
the $E_2(I)-E_1(I)$ varies from 100 to 250 keV, while at
$\gamma=15^\circ$ and $45^\circ$, the difference is around 450 keV.

In the right panel of Fig.~\ref{fig:specgam}, the results for a pure
$h_{11/2}$ proton particle and a neutron quasiparticle with
$\lambda_n=\varepsilon_5$ and $\Delta=1$ MeV are shown. The
$E_2(I)-E_1(I)$ curves are still parabola-like, while their minima
change with the spin $I$. The tendency is that the $\gamma$
deformation with the minimum energy difference decreases with spin.
It is noted that for $\gamma \in (20^\circ, 30^\circ)$, a near
constant energy difference ($\sim$ 200-250 keV) is observed. Also,
the energy difference between the yrast and yrare bands is quite
large (exceeds 450 keV) when the nuclear triaxiality is not
prominent, i.e., $\gamma\leq 15^\circ$ or $\gamma\geq 45^\circ$.

Among the candidate chiral doublet bands observed experimentally,
there are cases with a degeneracy point, e.g., $^{134}$Pr with $\pi
h_{11/2}\otimes\nu h_{11/2}$~\cite{Petrache06}, $^{104}$Rh with $\pi
g_{9/2}\otimes\nu h_{11/2}$~\cite{Vaman04}, or cases with a near
constant energy difference, e.g.,
$^{126,~128,~130,~132}$Cs~\cite{Koike03, WangSY06a} and
$^{106}$Rh~\cite{Joshi04}. Ref.~\cite{Koike03} suggests that the
near constant energy difference may come from a deviation of the
core shape from maximum triaxiality and a less favorable treatment
for the valence proton and neutron as a particle-hole configuration.
Here our calculations show that either a deviation of the core shape
from maximum triaxiality or a deviation of the Fermi energy surface
from a particle-hole configuration will hinder the level degeneracy
and prefer a near constant energy difference.

It is also demonstrated that the small energy difference between the
doublet bands suggests a triaxiality $(20^\circ < \gamma <
40^\circ)$ for the nucleus, in comparison with a difference of more
than 450 keV for $\gamma\leq 15^\circ$ or $\gamma\geq 45^\circ$.

\subsection{Electromagnetic Properties}

Electromagnetic transition probabilities are critical observables
which carry important information on the nuclear intrinsic
structure. Using a simple model for a special configuration in
triaxial odd-odd nuclei, Koike et al. suggested the selection rules
for electromagnetic transitions in chiral geometry~\cite{Koike04}.
The selection rules yield staggering of $B(M1)/B(E2)$ and
$B(M1)_{in}/B(M1)_{out}$ values for the partner band as a function
of spin $I$, where $B(M1)_{in}$ and $B(M1)_{out}$ refer to reduced
electromagnetic probabilities for intraband and interband $\Delta
I=1$ transitions, respectively. Such staggering behavior has been
regarded as a fingerprint for chirality in odd-odd triaxial nuclei,
and has been extensively used to support the declaration of chiral
doublet bands~\cite{Vaman04}. It is also acknowledged that in ideal
chiral doublet bands the electromagnetic transition probabilities
must be identical or, in practice, very similar~\cite{Petrache06}.
In the following, the electromagnetic transition probabilities will
be investigated with two quasiparticles coupled with the triaxial
rotor model in order to study whether such behavior of the
electromagnetic transition probabilities will be influenced by
variations in configurations and triaxial deformation.

\begin{center}
 \begin{figure*}[h!]
  \centering
  \includegraphics[width=8cm]{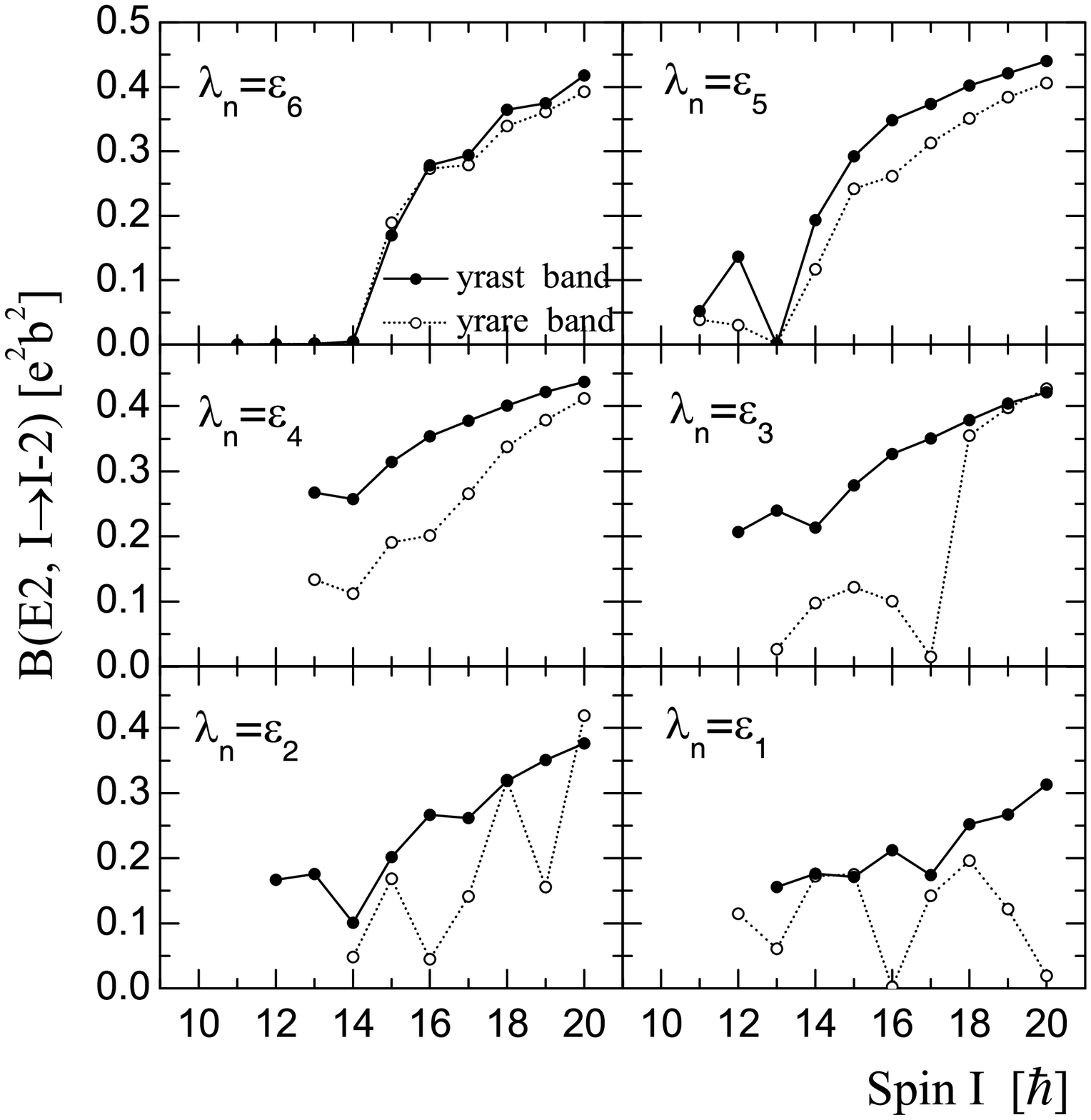}
  \includegraphics[width=8cm]{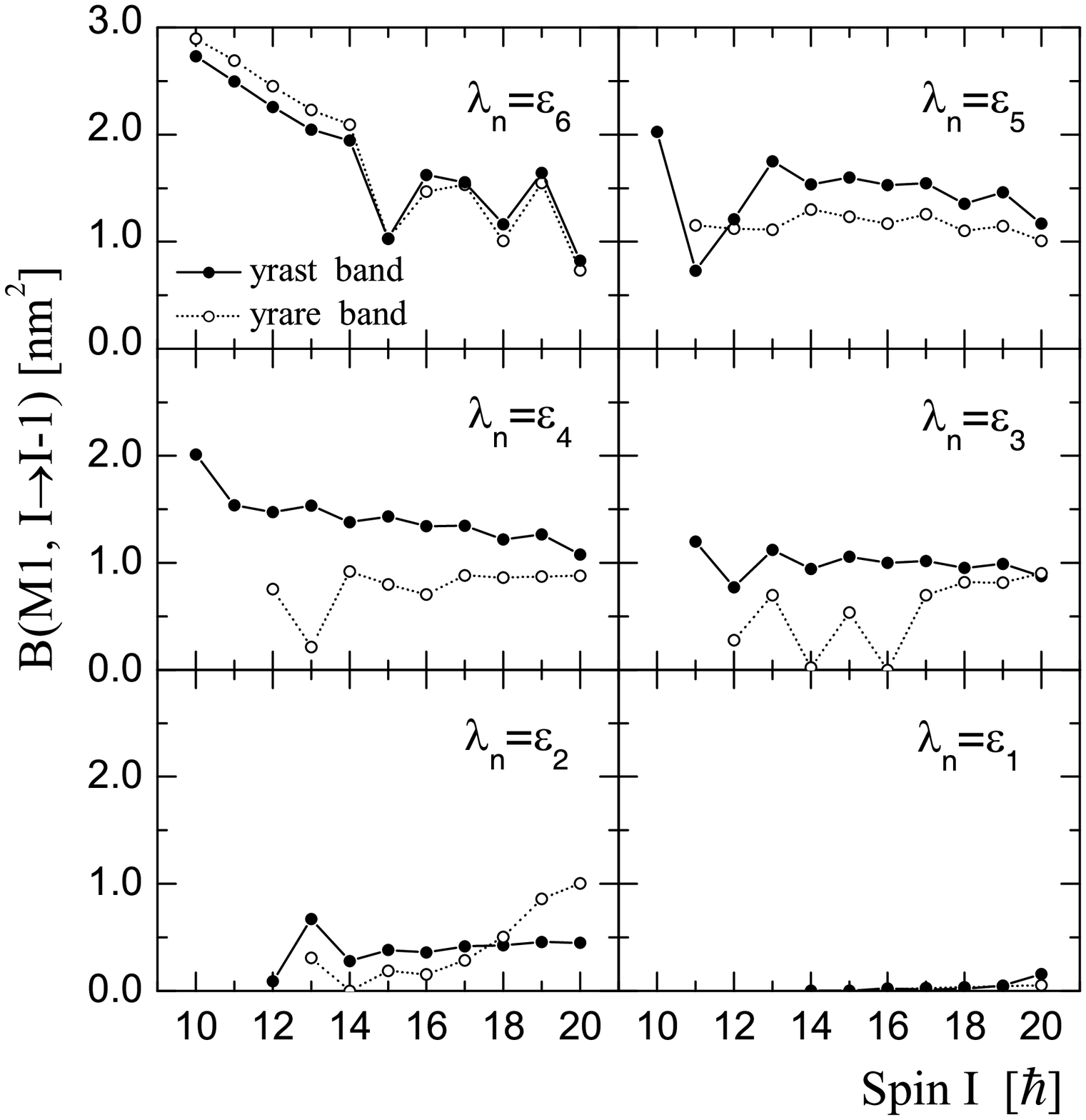}
   \caption{Calculated $B(E2)$ and $B(M1)$ values
     for the yrast and yrare bands:
     the same parameters as Fig. 2 are used.
   }
   \label{fig:BE2BM1_lamd}
\end{figure*}
\end{center}

Fig.~\ref{fig:BE2BM1_lamd} shows the intraband $B(E2)$ and $B(M1)$
values of yrast and yrare bands for different neutron Fermi energies
with $\gamma=30^\circ$. In the left panel, when
$\lambda_n=\varepsilon_{6}$, the intraband $B(E2)$ values at spins
$I\leq14$ are nearly zero. This is because the yrast and yrare bands
are displaced in energy for the lower spin region due to less
defined chiral geometry with insufficient collective rotation, and
these bands are mainly connected with $M1$ transitions. Note that
the interband $B(E2)$ values from yrare band to yrast band are large
in this spin region. For spin $I\geq15$, the intraband $B(E2)$
values increase gradually. For $\lambda_n=\varepsilon_{5}$, the
behavior of intraband $B(E2)$ is similar to the case
$\lambda_n=\varepsilon_{6}$, which is small at low spins, then
increases with spin. When $\lambda_n=\varepsilon_{4}$, or
$\varepsilon_{3}$, the intraband $B(E2)$ values of the yrare bands
have large differences in comparison with those of the yrast bands.
In general, the $B(E2)$ values of the yrast bands are larger than
those of the yrare bands, especially for spin $I\leq 17$. Their
values become close to each other with $I \geq 18$, where the
collective rotation of the deformed core makes an important
contribution to the total spin. For $\lambda_n=\varepsilon_{2}$, or
$\varepsilon_{1}$, the intraband $B(E2)$ values of the yrast band
increase with spin regularly, whereas those of the yrare band
exhibit many irregular oscillations.

For the intraband $B(M1)$ in the right panel of
Fig.~\ref{fig:BE2BM1_lamd}, the values of $B(M1)$ systematically
reduce as the neutron Fermi energy surface $\lambda_n$ decreases
from $\varepsilon_{6}$ to $\varepsilon_{1}$. When
$\lambda_n=\varepsilon_{1}$, the $M1$ transition almost vanishes
because both the valence proton and neutron are particle-like and
their contribution to the magnetic moment is canceled by similar
angular momentum orientations and different $g$-factor signs.
Therefore the rotation bands for $\lambda_n=\varepsilon_{1}$ are
mainly connected by $E2$ transitions, and correspond to the
so-called doubly decoupled bands. For $\lambda_n=\varepsilon_{6}$
and $\varepsilon_{5}$, the intraband $B(M1)$ values of yrast and
yrare bands are similar to each other. It can also be seen that the
odd-even staggering of $B(M1)$ for $\gamma=30^{\circ}$ is obvious
when $\lambda_n=\varepsilon_{6}$, while not so obvious in other
cases.

\begin{center}
 \begin{figure*}[h!]
  \centering
   \includegraphics[width=8cm]{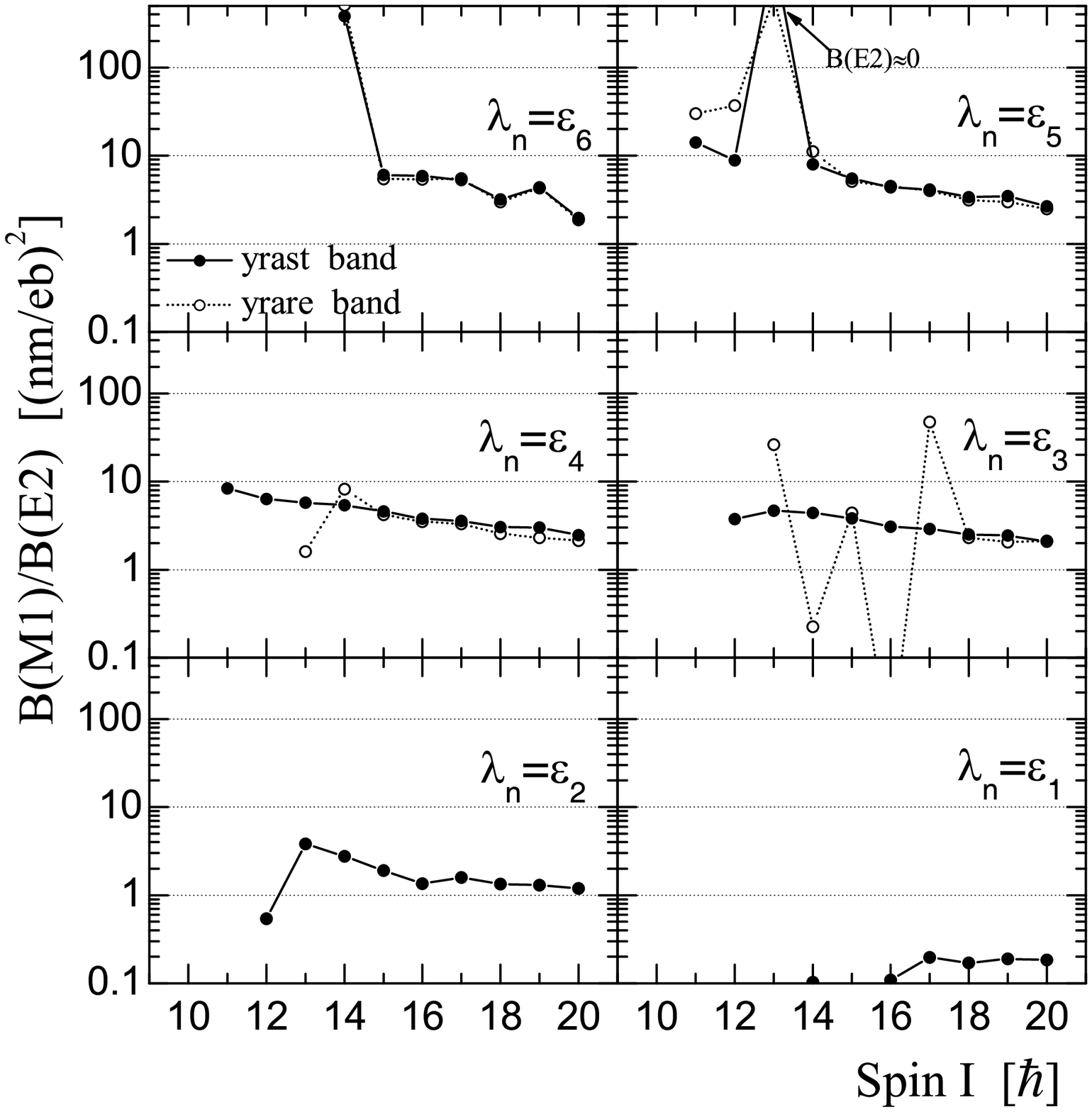}
   \caption{Calculated $B(M1)/B(E2)$ values
     for the yrast and yrare bands: the same parameters as Fig. 2 are used.
   }
   \label{fig:BM1E2_lamd}
\end{figure*}
\end{center}

Fig.~\ref{fig:BM1E2_lamd} shows the $B(M1)/B(E2)$ ratios of yrast
and yrare bands for different $\lambda_n$ with $\gamma=30^\circ$,
whereas the ratios of ``yrare bands" for $\lambda_n=\varepsilon_{2}$
and $\varepsilon_{1}$ are not presented due to their irregular
$B(E2)$ values.  It is interesting to note that the $B(M1)/B(E2)$
values in partner bands are close to each other for
$\lambda_n=\varepsilon_6$, $\varepsilon_5$, and $\varepsilon_4$, in
particular for higher spins, although there are noticeable
differences respectively in $B(E2)$ and $B(M1)$ values in
Fig.~\ref{fig:BE2BM1_lamd}. Next we examine the odd-even staggering
of $B(M1)/B(E2)$ ratios. For $\lambda_n=\varepsilon_6$, staggering
can be found for $I>16$ in the partner bands due to the staggering
of $B(M1)$ values. For $\lambda_n=\varepsilon_5$, a delicate
staggering for $I>16$ can be also seen. Except for $I<18$ in the
yrare band for $\lambda_n=\varepsilon_3$, there is no staggering
behavior of the $B(M1)/B(E2)$ ratios in the other yrast and yrare
bands.

\begin{center}
 \begin{figure*}[h!]
  \centering
  \includegraphics[width=8cm]{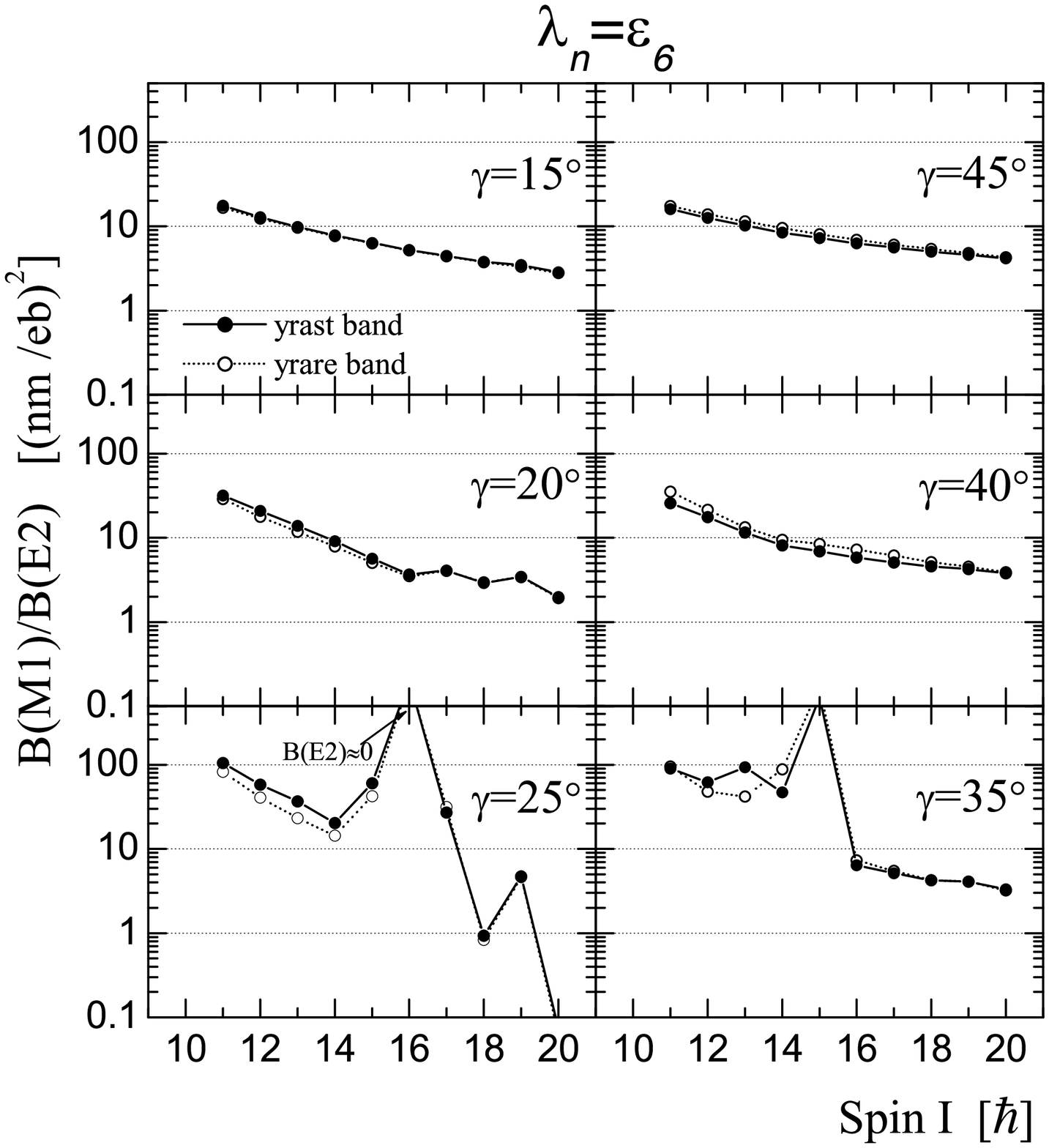}
  \includegraphics[width=8cm]{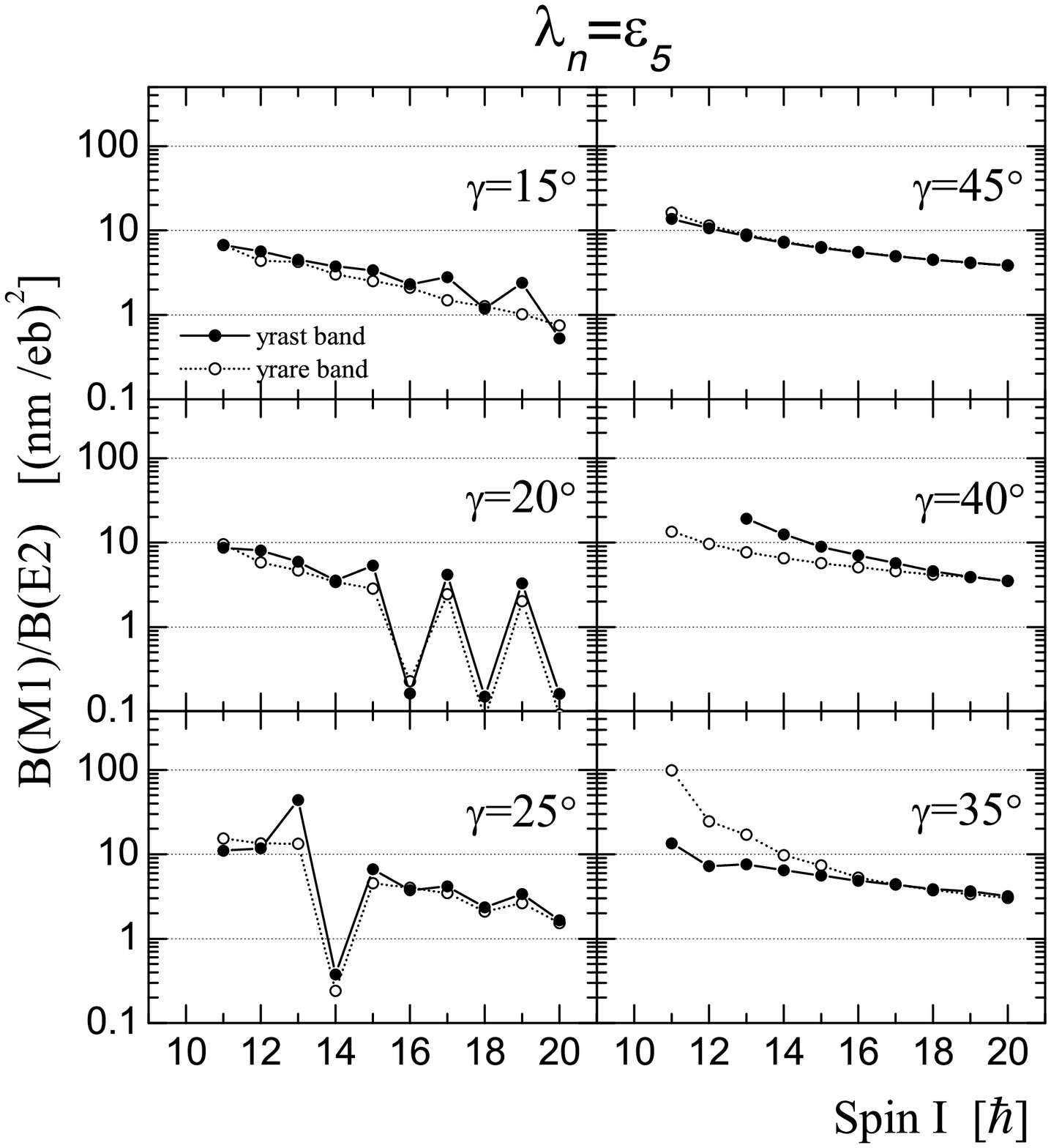}
   \caption{Calculated $B(M1)/B(E2)$ values for the yrast and yrare bands:
     the same parameters as Fig. 3 are
     used; $\lambda_n=\varepsilon_6$ (Left panel), $\varepsilon_5$ (Right
     panel).
   }
   \label{fig:BM1E2_gamma}
\end{figure*}
\end{center}

Fig.~\ref{fig:BM1E2_gamma} shows that the $B(M1)/B(E2)$ values for
the yrast and the yrare bands at different $\gamma$, i.e.,
$\gamma=15^{\circ}, 20^{\circ}, 25^{\circ}, 35^{\circ}, 40^{\circ},
45^{\circ}$, with neutron Fermi energy $\lambda_n=\varepsilon_6$
(left panel) and $\lambda_n= \varepsilon_5$ (right panel),
respectively. The results with $\gamma=30^{\circ}$ have been
presented in Fig.~\ref{fig:BM1E2_lamd}. One finds that: (1) for all
$\gamma$ degrees, the values of $B(M1)/B(E2)$ for the yrast bands
are close to those in yrare ones not only for
$\lambda_n=\varepsilon_{6}$, but also for
$\lambda_n=\varepsilon_{5}$ (3 neutron holes approximately); (2) the
staggering of $B(M1)/B(E2)$ ratios sensitively depends
on the deformation $\gamma$.

\subsection{Orientations of Angular Momenta}

The key to the formation of chiral bands in triaxial nuclei is the
existence of aplanar total angular momentum. Using wave functions
obtained from the particle-rotor model, one can calculate the
expectation values of angular momenta,  $\langle
\hat{I}_{i}\rangle$, $\langle \hat{j}_{i}\rangle$, and $\langle
\hat{R}_{i}\rangle$. The expectation values for the three components
of the angular momenta $\vec{I}, \vec{R}$, and $\vec{j}_p,
\vec{j}_n$ are given as,
 \beqn
&&\bar{I_i}\equiv \sqrt{\langle \hat{I}_i^2 \rangle},\nonumber\\
&&\bar{j_i}\equiv \sqrt{\langle \hat{j}_i^2 \rangle}, \nonumber\\
&&\bar{R_i}\equiv \sqrt{\langle (\hat{I}_i- \hat{j}_i)^2 \rangle}.
 \eeqn

\begin{center}
 \begin{figure*}[h!]
  \centering
  \includegraphics[width=8.0cm]{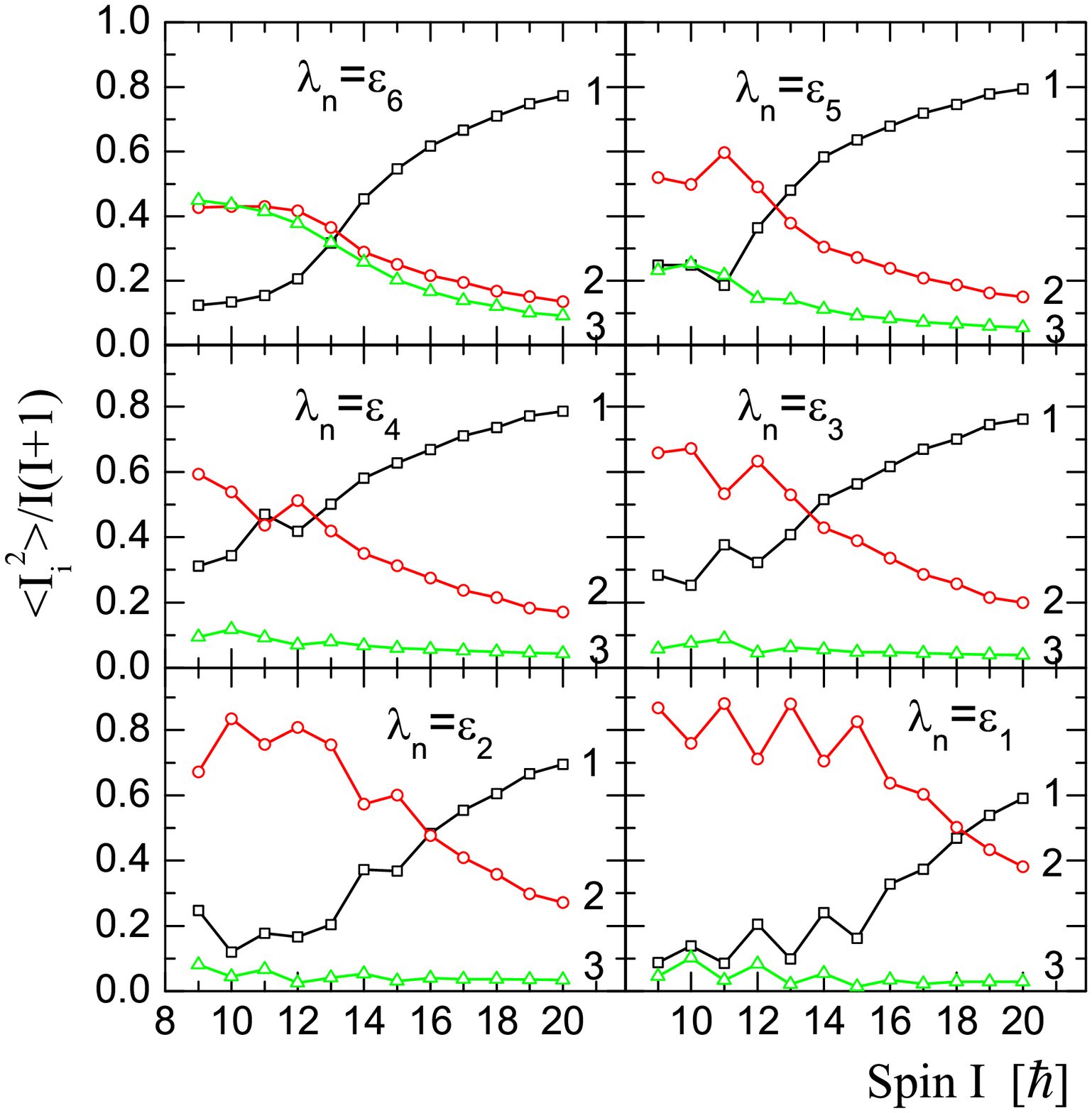}
  \includegraphics[width=8.0cm]{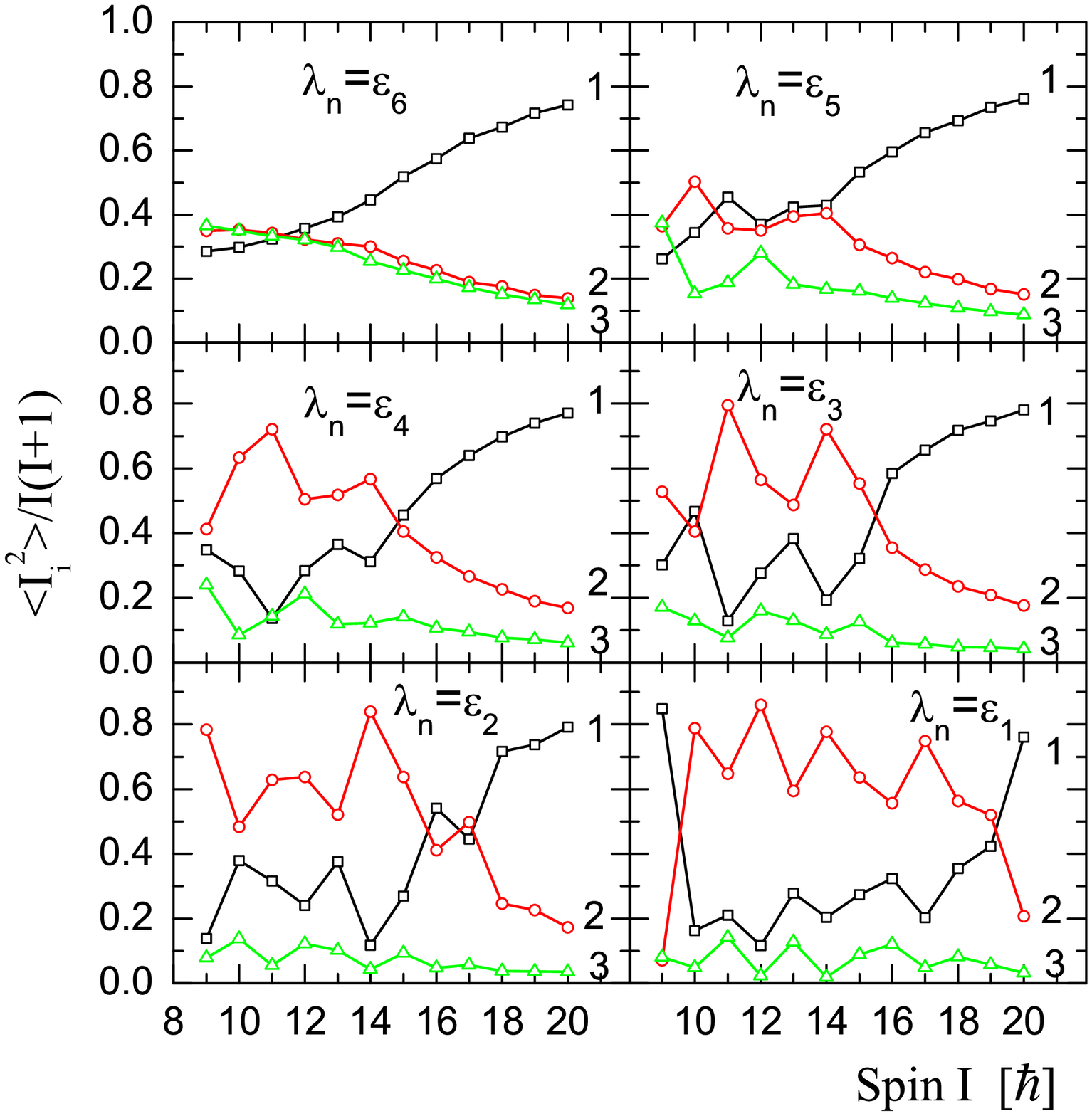}
   \caption{(Color online).
   For the yrast and yrare bands,
   the average contribution of three components to the total
   angular momentum $\langle \hat{I}_{i}^2\rangle/ I(I+1), i=1, 2, 3$
   in the intrinsic frame is plotted as a function of spin $I$:
   the same parameters as Fig. 2 are used.
   Open squares: 1-axis, open circles: 2-axis, open triangles:
   3-axis.}
   \label{fig:totang}

\end{figure*}
\end{center}

In Fig.~\ref{fig:totang}, the average contributions of the three
components $\bar{I_{i}}^2/I (I+1),~i=1,2,3$ to the total angular
momentum, are plotted for the yrast band (left panel) and yrare band
(right panel) with $\lambda_n$ changing from $\varepsilon_ {6}$ to
$\varepsilon_ {1}$. In the calculations, $\gamma =30^{\circ}$,
1-axis refers to the intermediate axis with the largest moment of
inertia, and the 2-, and 3-axis are respectively the short and the
long axis with ${\cal J}_2={\cal J}_3=1/4{\cal J}_1$. In all panels,
it can be seen that the average contributions from $I_1$ increase
globally with the total spin, while contributions from the other two
directions decrease globally.

For $\lambda_n=\varepsilon_{6}$, around $I=13$, the contributions
from three directions are comparable for both yrast and yrare bands.
This corresponds to a typical case of aplanar rotation. In fact, the
contributions to the total angular momentum from all three
directions are not negligible\footnote{According to quantum physics,
the minimum contribution from one direction to an angular momentum
$I$ is given by the value $\{\frac {1} {2} \ls I (I+1) -I^2\rs \}/I
(I+1)$, when the angular momentum is perpendicular to this
direction.} in the spin interval ($9<I<20$). Therefore the aplanar
solution is realized for this spin interval and chiral doublet bands
are expected. The statement is also true for the case
$\lambda_n=\varepsilon_5$, with the exception that the contribution
from the 3rd component is a little smaller compared with the case
$\lambda_n=\varepsilon_{6}$. As the Fermi energy surface $\lambda_n$
decreases, the contribution from the 3rd component becomes smaller,
the total angular momentum will mainly lie in the 1-2 plane, and an
aplanar rotation of the nucleus becomes a planar one. For both
$\lambda_n=\varepsilon_{4}$ and $\varepsilon_{3}$, aplanar solutions
can only be expected around $I \sim 11$. For
$\lambda_n=\varepsilon_{2}$ and $\varepsilon_{1}$, there exist only
planar rotations. In Fig.~\ref{fig:totang}, there are some
fluctuations of $\bar{I_{i}}^2/I (I+1)$ for
$\lambda_n=\varepsilon_1$, $\varepsilon_2$, $\varepsilon_3$, and
$\varepsilon_4$, due to the strong interactions between different
bands.

\begin{center}
 \begin{figure*}[h!]
  \centering
  \includegraphics[width=8cm]{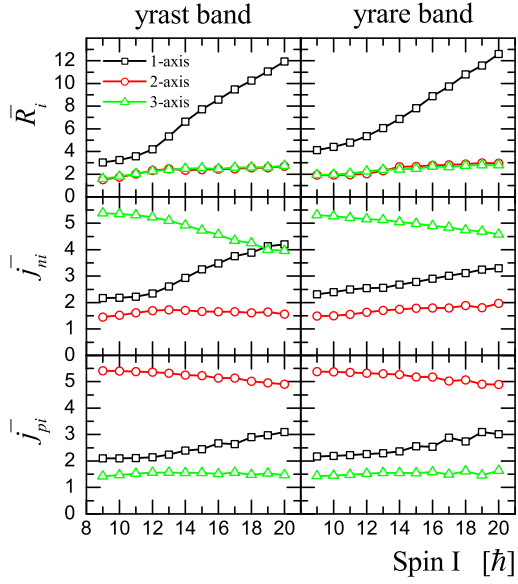}
   \caption{(Color online). For the yrast and yrare bands,
   the expectation values for the three components of the collective,
   odd-neutron, and odd proton angular momenta ---
   defined by $\bar{R_i} = \sqrt{\langle \hat{R}_{i}^2\rangle}$,
   $\bar{j_{pi}} =\sqrt{ \langle \hat{j}_{pi}^2\rangle}$, and
   $\bar{j_{ni}} = \sqrt{\langle \hat{j}_{ni}^2\rangle}$, (i=1, 2, 3) ---
   are plotted as functions of spin $I$:
   the same parameters as Fig. 2 are used, except that $\lambda_n=\varepsilon_6$.
   Open squares, open circles and open triangles correspond to the
   1-axis, 2-axis and 3-axis, respectively.}
   \label{fig:ang_e6}
\end{figure*}
\end{center}

\begin{center}
 \begin{figure*}[h!]
  \centering
  \includegraphics[width=8cm]{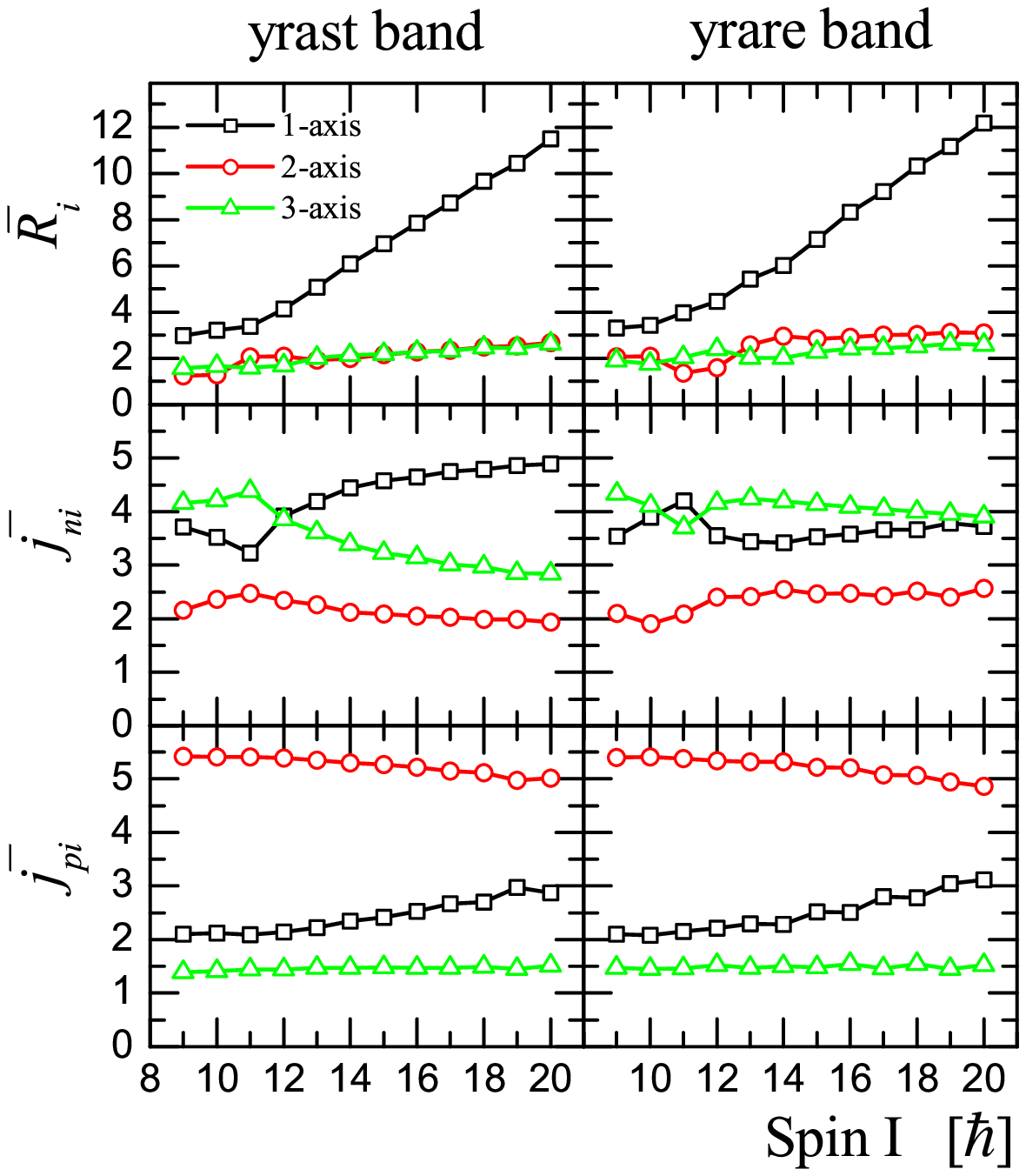}
   \caption{Same as Fig.\ref{fig:ang_e6}, except that
   $\lambda_n=\varepsilon_5$.
   }
   \label{fig:ang_e5}
\end{figure*}
\end{center}

\begin{center}
 \begin{figure*}[h!]
  \centering
  \includegraphics[width=8cm]{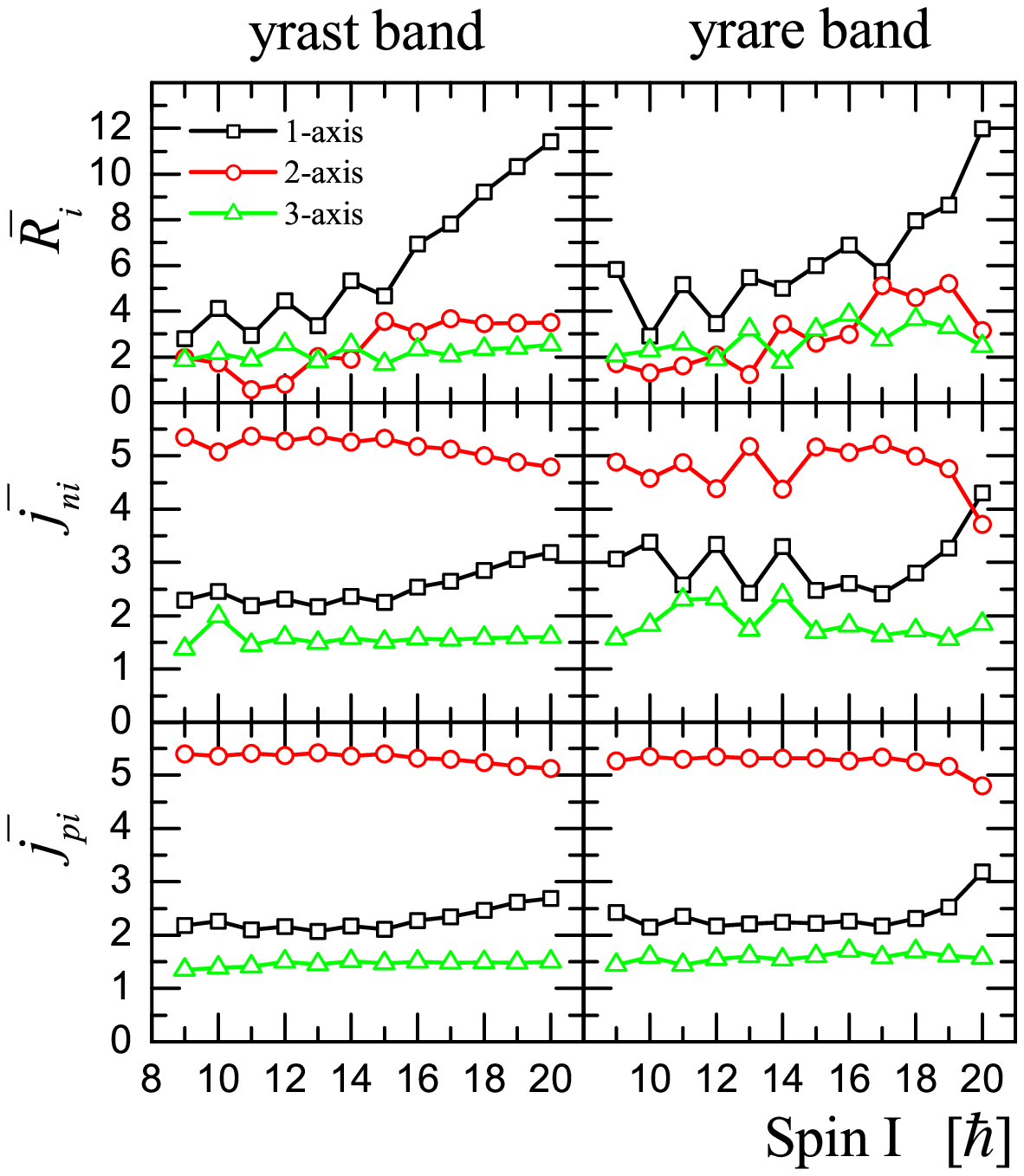}
   \caption{Same as Fig.\ref{fig:ang_e6}, except that
   $\lambda_n=\varepsilon_1$.
   }
   \label{fig:ang_e1}
\end{figure*}
\end{center}

The expectation values $\bar{R_i}$, and $\bar{j_{pi}}, \bar{j_{ni}}$
have been investigated for $\lambda_n = \varepsilon_6, \cdots,
\varepsilon_1$ as functions of the spin $I$. For simplicity, the
cases for $\lambda_n = \varepsilon_6$, $\varepsilon_5$, and
$\varepsilon_1$ are shown in Figs.~\ref{fig:ang_e6},
\ref{fig:ang_e5} and \ref{fig:ang_e1}, respectively.

In Fig.~\ref{fig:ang_e6}, for $\lambda_n = \varepsilon_6$, similar
as in Refs.~\cite{FM97,Starosta01}, the collective angular momentum,
and valence-proton and -neutron angular momentum align along the
intermediate axis (1-), the short axis (2-) and the long axis (3-)
respectively. Since these three angular momenta are mutually
perpendicular, a chiral picture results. In Fig.~\ref{fig:ang_e5},
for $\lambda_n = \varepsilon_5$,  the configuration is similar to
one proton plus three neutron holes in a single $h_{11/2}$ shell. In
this case, the orientations of $\vec R$ and ${\vec j}_p$ are similar
to Fig.~\ref{fig:ang_e6}, while the third component of the angular
momentum ${\vec j}_n$ is reduced. The total angular momentum
$\vec{I}$ is still aplanar, but its inclination angle to the 1-2
plane becomes smaller. As the neutron Fermi energy surface
decreases, the hole-like odd neutron will switch to a particle-like
one, and $\vec{j}_n$ will align from the 3-axis to the 2-axis. Then
the valence proton and neutron both align to the 2-axis, with the
collective angular momentum along the 1-axis, and together they give
the total angular momentum in the 1-2 plane. This is a planar
solution as shown in Fig.~\ref{fig:ang_e1}. Noted that in all cases
the expectation values along the 1-axis for $\vec{j}_n$ and
$\vec{j}_p$ increase with $I$ due to the rotational alignment of odd
particles.

The average core contribution to the total angular momentum can be
seen in the upper panels of Figs.~\ref{fig:ang_e6},
\ref{fig:ang_e5}, \ref{fig:ang_e1}. In Fig.~\ref{fig:ang_e6}, we
note that the core contribution for the $14^+$ state in both the
yrast and the yrare band ($R\sim 6.5\hbar$) is comparable with the
contributions from the valence proton and valence neutron. The
latter is consistent with the result in Ref.~\cite{Starosta01npa}.
In Figs.~\ref{fig:ang_e6} and \ref{fig:ang_e5}, $\bar{R_1}$
increases by 8$\hbar$ (from $\sim4$ to $\sim12\hbar$) as the spin
$I$ changes from $12\hbar$ to $20\hbar$. This demonstrates that the
increase of the total angular momentum is mainly due to the
collective rotation for $I\geq 12$. Therefore the transition
probabilities $B(E2, I\rightarrow I-2)$ corresponding to the
collective rotation should be large for $I\geq14$. These results are
consistent with the $B(E2)$ values discussed in
Fig.~\ref{fig:BE2BM1_lamd}. For the lower spin region near the
bandhead ($I\leq12$), the increase of $I$ comes mainly from the
contributions of the valence proton and neutron while the
contribution from the core stays the same. For
$\lambda_n=\varepsilon_1$ in Fig.~\ref{fig:ang_e1}, the collective
angular momentum at low spin range $I<16$ exhibits odd-even
staggering which is consistent with the energy spectrum in
Fig.~\ref{fig:spenergy}.

\section{Conclusion}

A particle rotor model with a quasi-proton and a quasi-neutron
coupled with a triaxial rotor is developed and applied to study
chiral doublet bands with configurations of a $h_{11/2}$ proton and
a $h_{11/2}$ quasi-neutron. With pairing correlations taken into
account by the BCS method, a proton and many neutron holes coupled
with a triaxial rotor can be simulated by changing the neutron Fermi
level from the top $h_{11/2}$ orbit $\varepsilon_6$ to the lowest
one $\varepsilon_1$.

The energy spectra, electromagnetic properties, as well as the
orientations of the angular momenta of the doublet bands have been
investigated in detail. The results are summarized as follows:
 \begin{enumerate}
 \item Aplanar rotation exists at least for $\lambda_n=\varepsilon_{6}$
       and $\lambda_n=\varepsilon_5$ in a certain spin interval. The
       contributions from the three axes are comparable to each other for
       the partner bands. This demonstrates that chiral geometry
       holds even for the valence nucleons deviating from a pure
       particle-hole configuration.
 \item The near constant energy separation ( $\sim 200$ keV )
       between the partner bands, which has been observed in many
       candidate chiral bands experimentally,
       has been obtained for $\lambda_n=\varepsilon_6$ and
       $\lambda_n= \varepsilon_5$ for certain spin and
       deformation $\gamma$ intervals.
 \item Either a deviation of the core shape from $\gamma=30^{\circ}$ or
       a deviation of the Fermi energy surface from a particle-hole
       configuration will hinder the level degeneracy and prefer a near
       constant energy difference.
 \item For $15^\circ\leq\gamma\leq45^\circ$, $\lambda_n$ lies between $\varepsilon_{6}$
       and $\varepsilon_{5}$, the $B(M1)/B(E2)$ values
       together with $B(E2)$ , $B(M1)$ for the yrast bands are close
       to those in yrare bands, which may hold for all chiral bands.
 \item The odd-even staggering of $B(M1)/B(E2)$ values
       is strongly influenced by the deformation $\gamma$
       as well as the Fermi surface $\lambda$, which suggest
       that the odd-even staggering
       of $B(M1)/B(E2)$ values may not be a general feature
       for the chiral bands.
 \end{enumerate}

With pairing treated by the BCS approximation, the present
quasi-particles PRM is aimed at simulating one proton and many
neutron holes coupled with a triaxial rotor. After a detailed
analysis of the angular momentum orientations, energy separation
between the partner bands, and behavior of electromagnetic
transitions, it is demonstrated that aplanar rotation or
equivalently chiral geometry, does exist beyond the simple one
proton and one neutron hole coupled with a triaxial rotor. While
simulating multiple valence particles here by adjusting the Fermi
energy, one may argue that the valence particles dumped into the BCS
vacuum in the present model can not contribute to the moments of
inertia. However, as the main focus is the nuclei in the A $\geq$
100 mass region, the influence by such approximation should not
result in a serious problem.  Of course, a model with multi-proton
particles (holes) and multi-neutron holes (particles) coupled
explicitly with a triaxial rotor is necessary. Future work should
also be devoted to replace the present single-$j$ shell by a more
realistic single particle potential, such as, Nilsson potential.

\begin{acknowledgments}
The authors express their sincere gratitude to G.C.~Hillhouse for
helpful comments regarding this manuscript. This work is supported
by the National Natural Science Foundation of China under Grant
Numbers 10505002, 10435010, 10605001, and 10221003, the Postdoctoral
Science Foundation of China under Grant No. 20060390371.
\end{acknowledgments}

\end{document}